\documentclass[referee]{raa}
\usepackage{graphicx,times,amsmath,amssymb,natbib}             
\usepackage{natbib}
\bibpunct{(}{)}{;}{a}{}{,}
\usepackage{color}
\usepackage{subfigure}
\usepackage{upgreek}

\usepackage[a4paper=true,dvipdfm=true,pagebackref=true]{hyperref}
\hypersetup{pdftitle = The title of my PDF, pdfauthor = My name, pdfsubject= The subject, pdfkeywords = keyword1 keyword2 keyword3} 
\hypersetup{colorlinks = true, linkcolor = green, anchorcolor = red, citecolor = blue, filecolor = red, pagecolor = red, urlcolor = red}
\begin{document}
\title{AstroSat observation of GX 5-1: Spectral and timing evolution}
 \volnopage{Vol.0 (200x) No.0, 000--000}      
	\setcounter{page}{1}          
	
\author{Yashpal Bhulla\inst{1} \and Ranjeev Misra\inst{2} \and J.S. Yadav\inst{3} \and S.N.A Jaaffrey\inst{1} } 
\institute{ Pacific Academy of Higher Education and Research University, Udaipur-313003, India \\
\and Inter-University Center for Astronomy and Astrophysics, Pune-411007, India\\
\and Tata Institute of Fundamental Research, Mumbai-400005, India \\
}

\abstract{
We report on the first analysis of AstroSat observation of the Z-source GX 5-1 on February 26-27, 2017. The hardness-intensity plot reveals that the source traced out the horizontal and normal branches. The 0.8-20 keV spectra from simultaneous SXT and LAXPC data at different locations of the hardness-intensity plot can be well described by a disk emission and a thermal Comptonized component. The ratio of the disk flux to the total i.e. the disk flux ratio increases monotonically along the horizontal to the normal one. Thus, the difference between the normal and horizontal branches is that in the normal branch,  the disk dominates the flux  while in the horizontal one it is the Comptonized component which dominates. The disk flux scales with the inner disk temperature as $T_{in}^{5.5}$ and not as $T_{in}^4$, suggesting that either the inner radii changes dramatically or that the disk is irradiated by the thermal component changing its hardness factor. The power spectra reveal a Quasi Periodic Oscillation whose frequency changes from $\sim$ 30 Hz to 50 Hz. The frequency is found to correlate well with the disk flux ratio. In the 3-20 keV LAXPC band the r.m.s of the QPO increases with energy (r.m.s $\propto$ $E^{0.8}$), while the harder X-ray seems to lag the soft ones with a time-delay of a milliseconds. The results suggest that the spectral properties of the source are characterized by the disk flux ratio and that the QPO has its origin in the corona producing the thermal Comptonized component.
}
\authorrunning{Bhulla et al}
\titlerunning{AstroSat observation of GX 5-1}

 \maketitle
\section{Introduction}
X-ray Binaries are the class of binaries that emit X-rays radiations. An X-ray Binary system has a normal star or a white dwarf
transferring mass onto a compact object such as a neutron star (NS) or a Black Hole (BH). These systems can be classified into two categories, i.e., low-mass X-ray binaries (LMXB) and high-mass X-ray binaries (HMXB), based on the mass of the companion star \citep{1989hasi}. In LMXBs the low mass companion star is usually a late type star and matter accretes through Roche lobe overflow e.g. \citep{1999konar,
  2011paulb}. 

LMXB harbouring neutron stars can be characterized
into two classes called ``Atoll'' and ``Z'' based on the shapes they
track out in their  X-ray hardness intensity diagram (HID)[e.g. Atoll
  type : 4U 1608-52 \citep{2015leiy}, GX 3+1 \citep{2000kulk}, 4U
  1735-44 \citep{2013leiyj}, 4U 1728-34
    \citep{2003miglia}, 4U 1820-30 \citep{2017trigo}, GX 9+1
    \citep{2016wangj} \& Z type : Cyg X-2 \citep{2013leiyj}, GX 17+2
  \citep{2002kulk}, GX 349+2 \citep{2004iaria} Sco X-1
    \citep{2001titar}, GX 340+0 \citep{2013seif}, GX 5-1
    \citep{2009jackson}] \citep{1989hasi}.  The HID is a
plot of hard colour  as a function of total X-ray count rates. The HID
plot representations are useful to understand the variations in
spectra and the spectral components. 

 The Z-type sources are characterised
  by having three main branches in a HID. From top-left to
bottom-right is the horizontal branch (HB), and then the  normal
branch (NB) and the bottom flaring branch (FB). The transition between
the NB and HB is called the Hard Apex (HA) and between the FB and NB
is called Soft Apex \citep{1993blom, 2003agrawal}. Z-type sources are
further subdivided into two subcategories: Those with a somewhat
vertical or short HB ,  but with an extended long FB are called ``Sco
like sources'', (e.g., Sco X-1, GX 17+2 and GX 349+2) while those
sources that have a proper horizontal HB and weak FB are called ``Cyg
like sources'', (e.g., Cyg X-2, GX 340+0, GX
5-1)\citep{1991penn,1994kuul,2012church}. 

Theoretically, for NS-LMXBs the X-ray emission may originate from the
accretion disk and a boundary layer between the disk and the neutron
star surface
\citep{1989hanawa,2001popham,2003gilfanov}. Observationally, the
spectra of these sources are dominated by a Comptonized component and
a soft one. There has been controversy regarding the origin of these
components. In one interpretation, the soft component is due to a
multi-colored disk and the Comptonized one due to the boundary layer
\citep{1984mitsuda, 1986makishima,1989mitsuda,1999barret,2009agmis} while the
alternative is that the soft component is a black body emission from
the boundary layer and the Comptonized one is due to a corona
\citep{2004church, 2006kong, 2008lavagetto}. In fact some spectral analysis suggest the more natural situation that the seed photons arise
  from two components (i.e. a black body and a disk emission) or that one
can dominate at different regions of the Z-track \citep{2011raichur,2012sakurai,2017padilla}. The spectral parameters
vary along the track but are subject to which
interpretation is chosen \citep{2006falanga,1988whites, 1998titar,
  2003agrawalshree}.  It has
  been argued that changes in the mass accretion rate, $\dot{M}$, are responsible for a source changing its spectral state or moving along its
branches in the HID, as proposed by \cite{1986pried} .
Support for an increase of $\dot{M}$ from the HB to the NB, and then
along the FB, came from a multi-wavelength study of Cyg
X-2 by Hasinger et al. (1990). In the past decade, however,
it has been shown that the variability of Z sources is not so
simple as this \citep{2012church}. For example, spectral analysis along the Z-track suggests that
the accretion rate maybe constant \citep{Lin09}.

Another debate in the literature has been the location
  and size of the region producing the Comptonizing component. For eclipsing
  sources, the time taken for the companion star to eclipse the X-ray
  producing region suggest a large size of $\sim 10^{11}$ cms \citep{2004church}
  and hence the spectra of these sources have been interpreted in terms
  of the extended accretion disk corona model \citep{2014churuch}. An extended
  hot region is also required to produce the ionized emission lines seen
  in high resoultion spectra \citep{2009schulz}. However it is not clear how
  the gravitational energy which is released close to the neutron star is
  transported to power in such  an extended corona. More importantly, as discussed
  below, these sources
  exhibit high frequency quasi periodic oscillations which demand a
highly compact origin of the Comptonized component instead of an extended one.

Timing studies are an important, possible way to constrain the
geometry and interpret the changes along the Z-track. These studies
are generally undertaken by computing the power density spectra (PDS)
which is the amplitude square of the Fourier transform of the source
light curve. Its generally consist of broad features along with narrow
peaks with the latter being called Quasi-Periodic Oscillations (QPOs)
\citep{1999mendez,2000vanderklis,2000mendez,2001mendez,2005barret,2018mond}. It
should be noted that most of these studies are based on analysis of
data from the Rossi X-ray Timing Explorer (RXTE). QPOs have been
observed in the horizontal and normal branches and are named as
horizontal branch oscillations (HBOs) and normal branch oscillations
(NBOs) respectively \citep{1985vander,1991lamb,1992lewin}. The
frequencies of the HBO lie in the range of 15 - 60 Hz while for NBO
they lie between 5 - 20 Hz \citep{2002jonker,2007homan}. It is also
known that the frequencies evolve along the track i.e. the HBO
frequency decreases from the hard apex as the source moves along the
horizontal branch and the frequency is correlated with the
intensity. These sources also show very rapid variability in the form
of kHz QPOs which are seen when the source in the normal and flaring
branches\citep{1996kulk,1998wijnanads}.

GX 5-1 is the second brightest (after Sco X-1, \citep{1962giacc})
persistent NS LMXB. The source is classified as Z-track LMXB and is
located near to the Galactic center. It radiates nearly at or
exceeding the Eddington limit with typical Eddington fraction
$L/L_{edd}$ = 1.6-2.3 assuming a distance of 9.0 kpc
\citep{2009jackson,2011srirama,2018homan}. It is a Cyg-like source
having an extended horizontal branch \citep{1994kuul}.The  source
exhibits NBO and HBO along with kHz QPOs \citep{1999wijnands,
  2002jonker}. The frequency of its HBO ranges from 13 to 50 Hz and is
known to correlated with X-ray luminosity \citep{1985vander}. Hard
time lags (i.e. the hard photons are delayed w.r.t to the soft ones)
of the order of milliseconds, have been detected for both HBOs and
NBOs \citep{1987vanderk, 1994vaughan, 2011srirama}.

RXTE spectral analysis of GX 5-1 was conducted by \citet{2009jackson}
, who favored a blackbody and Comptonized
components (which they refer to as an extended accretion
disk corona model) over one where the emission is
from a multi-color disk and Comptonization. However, \cite{2011srirama} found by analyzing RXTE data that the
  multi-color disk and Comptonization component provided an
  acceptable fit and found evidence for the variation in the
  inner disk radius. They also showed the r.m.s of the QPOs observed in
  both the
  normal and horizontal branches increase with energy suggesting a origin
  in the inner regions. A more extensive work involving a larger number of
  RXTE observations confirmed these basic results \citep{2012sriram}. However, we
note that such analysis were hampered by the absence of
 data below 3 keV. NuSTAR observation (again limited to energies above 3 keV) of GX 5-1 also
showed  a black-body component with temperature $\sim$2.5 keV at the
horizontal branch, which was interpreted as a boundary layer component
\citep{2018homan}.

AstroSat is the first Indian Multi-wavelength satellite. The Large
Area X-ray Proportional Counters (LAXPC) consists of 3
  identical proportionals counter units (PCUs) having a total effective
  area $\sim$ 6000 $cm^2$ at 15 keV and provides  event data list
which can be used effectively to compute PDS for different energy
bands \citep{2016yadava,2017agrawallaxpc,2017antia}. The LAXPC is well
complimented by the  Soft X-ray Telescope (SXT) providing simultaneous
coverage in the soft energy band \citep{2017singh}. The LAXPC has
already provided unprecedented data to study the rapid temporal
behaviour of black hole systems
\citep{2016yadavlaxpc,2017misra,2017bhargava,2018PAHARI}. It has
detected kHz QPOs and burst oscillation from the
neutron star system 4U 1728-34 \citep{2017chauhan}
and has provided detailed information on the source's thermo-nuclear
bursts \citep{2018BHATTACHARYA}. Spectral and timing properties of the
Atoll Source 4U 1705-44 along the banana track have been characterized
by LAXPC \citep{2018agrawal}. One of the primary advantage of
studying neutron star systems such as GX 5-1 by AstroSat is the
simultaneous broad band spectral data ($0.7-30$ keV provided by LAXPC and
SXT) with high resolution timing data. This provides an unique opportunity to
study the connection between the temporal (long and short term) properties
of the source with the spectral components.

In this work, we present results of AstroSat data analysis of the
Z-source GX 5-1. To the best of our knowledge this is the first time
that both LAXPC and SXT data have been used to track a neutron star
source as it evolves in its hardness diagram. The timing information
from LAXPC is combined with the wide band spectra of the LAXPC/SXT to
model the spectral and temporal evolution of this source. 

\begin{figure}
\centering \includegraphics[width=0.40\textwidth,angle=-90]{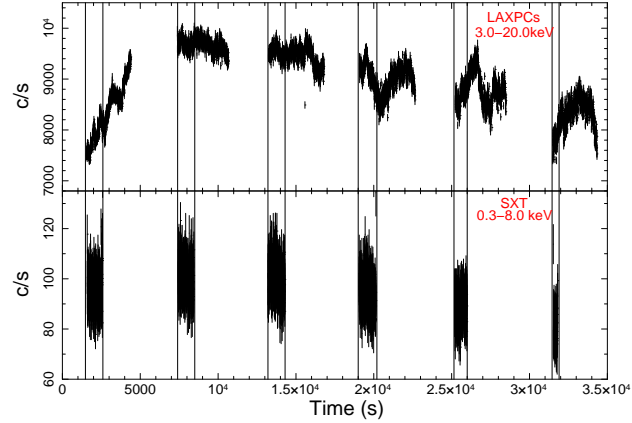}
\caption{Background subtracted light curve of the GX 5-1. The time bin is  2.3778 seconds at the
SXT time resolution. The vertical lines mark the regions from which strict simultaneous spectra were obtained. }
\label{msxxfig1}
\end{figure}
 
 \begin{figure}
\centering \includegraphics[width=0.40\textwidth,angle=-90]{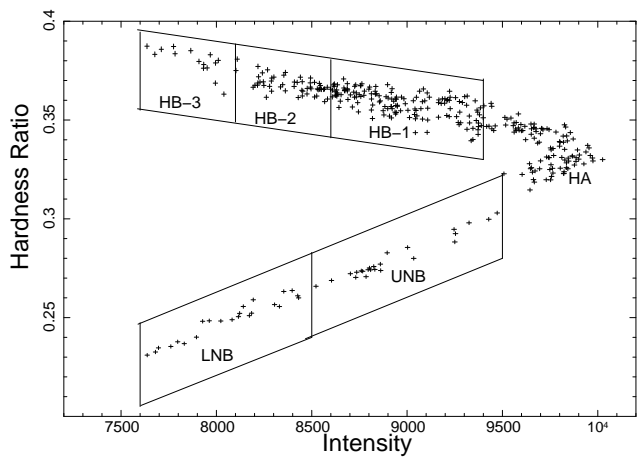}
\caption{Hardness Intensity Diagram (HID) of GX 5-1 observed with AstroSat/LAXPC. The horizontal and normal branches are clearly seen, which are subdivided into six regions for spectral analysis.}
\label{msxxfig2}
\end{figure}

\section{Observations and data analysis}

GX 5-1 was observed by AstroSat from February 26, 2017 17:34:56 till February 27, 2017 06:45:17 for an  effective exposure time of  $\sim$35 ks. The LAXPC data were analysed using the LAXPCsoftware (http://astrosat-ssc.iucaa.in/?q=laxpcData) to obtain background subtracted light curves, photon spectra, power density spectra and frequency dependent time-lags. To take into account LAXPC response uncertainties a systematic error of 3\% was used for spectral fitting. The LAXPC spectra were fitted from 3-20 keV  since the data was background dominated at higher energies. The SXT spectra and lightcurves were extracted using XSELECT (XSELECT V2.4d) from level-2 data. A circular region with a 12' radius was used for source extraction. Due to uncertainties in the effective area and response, the SXT spectra were considered in the energy range 0.8-6 keV and a gain fit correction was included. Simultaneously spectral fitting for both the instruments  were carried out using XSPEC version 12.9.1.

Figure \ref{msxxfig1} shows the LAXPC and SXT background lightcurves binned to 2.3778 seconds i.e at the SXT time resolution. The LAXPC lightcurve is the sum of all three detectors and is for the energy range 3-20 keV while for SXT it is 0.8-6 keV. Data gaps are due to SAA passages, earth occult and instrument shutdowns. SXT operation are at night time and hence its efficiency in viewing is less than that of LAXPC. The vertical lines mark the common time for the two instruments and as described in the next section, spectral analysis has been  done for strict simultaneous data.

We define hardness as the ratio  between the LAXPC count rates in 8-20 keV range divided by that in the 3-8 keV. The hardness was computed for 60 second time bins and is plotted against the total count rate (or intensity) in the 3-20 keV band in Figure \ref{msxxfig2}. In such a hardness intensity diagram (HID), portions of the Z-track are clearly distinguished. The source shows an extended horizontal branch, a hard apex and a normal branch. No flaring branch is seen for this observation. The normal branch corresponds to the start of the observation when in a time-scale of a few ksecs the source traced out the entire branch. For most of the observation the source was in the hard apex and horizontal branch.

For spectral and timing analysis we split the Z-track into six distinct regions. The horizontal branch was split into three regions named them HB-1, HB-2 and HB-3. The hard apex region was named HA and the normal branch was split into two: lower (LNB) and upper (UNB) normal branch. The selected regions for the horizontal and normal branches are shown as rectangular boxes in Figure \ref{msxxfig2}. The remaining part of the HID was marked as hard apex HA. The regions have been chosen to roughly span the Z-track and the effective exposure times for simultaneous LAXPC and SXT data of the
LNB, UNB, HA, HB-1,HB-2 and HB-3 are 1080, 2140 ,3557, 490, 484 and 304 seconds respectively.

 \begin{figure}
\centering \includegraphics[width=0.40\textwidth,angle=-90]{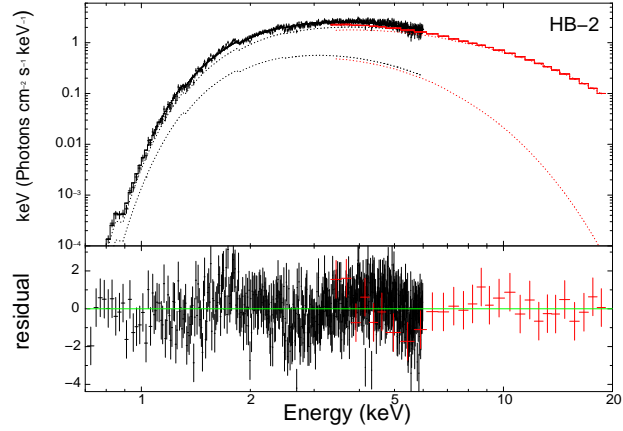}
\caption{The unfolded spectrum  0.8-6 keV SXT (black), 3-20 keV LAXPC in red spectrum of GX 5-1 for HB2 region.}
\label{msxxfig3}
\end{figure}

   \begin{table*}
   \begin{center}
   \begin{tabular} {c c c c c c c c c c}
 \hline 
 Location & $N_{H}$ & $N_{dbb}$ & $R_{in}$ & $kT_{in}$ & $\Gamma$  & $kT_{e}$ & $F_{T}$ & $F_{D}/F_{T}$ & $\chi^2_\nu/dof$ \\
  LNB &  $2.75_{-0.04}^{+0.04}$ &$226_{-20.22}^{+25.32}$ & $19.13_{-0.88}^{+1.04}$  & $1.40_{-0.12}^{+0.17}$ & $2.52_{-0.13}^{+0.09}$& $2.92_{-0.60}^{+1.60}$ & $4.13_{-0.03}^{+0.02}$ & $0.40_{-0.01}^{+0.01} $ & 1.35/523  \\ 
$^{**}$UNB & $2.75^f$ & $185^f$ & 17.31& $1.44_{-0.06}^{+0.05}$ & $2.68_{-0.07}^{+0.06}$ & $3.94_{-0.17}^{+0.21}$ & $3.41_{-0.04}^{+0.03}$ & $0.34_{-0.02}^{+0.01}$  & 0.82/24  \\
  HA & $2.74_{-0.01}^{+0.02}$ & $177_{-10.41}^{+11.10}$ & $16.93_{-0.51}^{+0.52}$ &  $1.46_{-0.05}^{+0.04}$ & $2.10_{-0.08}^{+0.07}$  & $2.88_{-0.04}^{+0.05}$  & $4.70_{-0.02}^{+0.02}$ & $0.32_{-0.01}^{+0.01}$ & 1.67/546 \\
  HB1 & $2.73_{-0.03}^{+0.03}$ & $183_{-39.48}^{+36.073}$ & $17.20_{-2.01}^{+1.63}$  & $1.30_{-0.14}^{+0.16}$  & $2.08_{-0.05}^{+0.05}$ & $2.85_{-0.27}^{+0.20}$  & $4.26_{-0.03}^{+0.04}$ & $0.23_{-0.01}^{+0.01}$ & 1.15/547  \\
  HB2 & $2.67_{-0.02}^{+0.06}$ & $119_{-45.71}^{+42.18}$  & $13.88_{-2.93}^{+2.26}$ & $1.26_{-0.15}^{+0.09}$  & $2.06_{-0.06}^{+0.05}$ & $2.97_{-0.16}^{+0.23}$ & $4.15_{-0.03}^{+0.05}$ & $0.13_{-0.01}^{+0.01}$ &  1.13/523  \\
  HB3 & $2.69_{-0.08}^{+0.10}$ & $98_{-16.45}^{+13.81}$ & $12.60_{-1.15}^{+0.85}$ & $0.96_{-0.09}^{+0.09}$  & $2.05_{-0.04}^{+0.03}$ & $2.95_{-0.05}^{+0.05}$  & $3.84_{-0.06}^{+0.08}$ & $0.03_{-0.01}^{+0.06}$ & 1.07/428 \\  

    \hline
    \end{tabular}
    \end{center}
   \caption{$N_{H}$ is absorption column density in units of $\times$ $10^{22}$ atom $cm^2$. $N_{dbb}$ = $\{R_{in}/D_{10}\}^2{\ast}cos$ $i$ is the normlisation of the disk blackbody component . $R_{in}$ is inner disk radius in km computed from $N_{dbb}$ assuming a distance of 9 kpc and inclination angle of $60^o$. $kT_{in}$ is the inner disk temperature.  $\Gamma$ and $kT_{e}$ are the photon index and electron temperature of the thermal Comptonization component.  $kT_{in}$ \& $kT_{e}$ are in keV unit. $F_{T}$  \& $F_{D}$ are the unabsorbed bolometric flux and the flux of disk component with unit erg $cm^{-2}$ $s^{-1}$. ($^f$) denotes that the parameter was frozen,  ($^{**}$) indicates that at this location there was no SXT observation and hence spectral fitting is done with LAXPC data alone.}
    \label{msxxtable1}
    \end{table*}

\begin{figure*}
  \centering \includegraphics[width=0.30\textwidth,angle=-90]{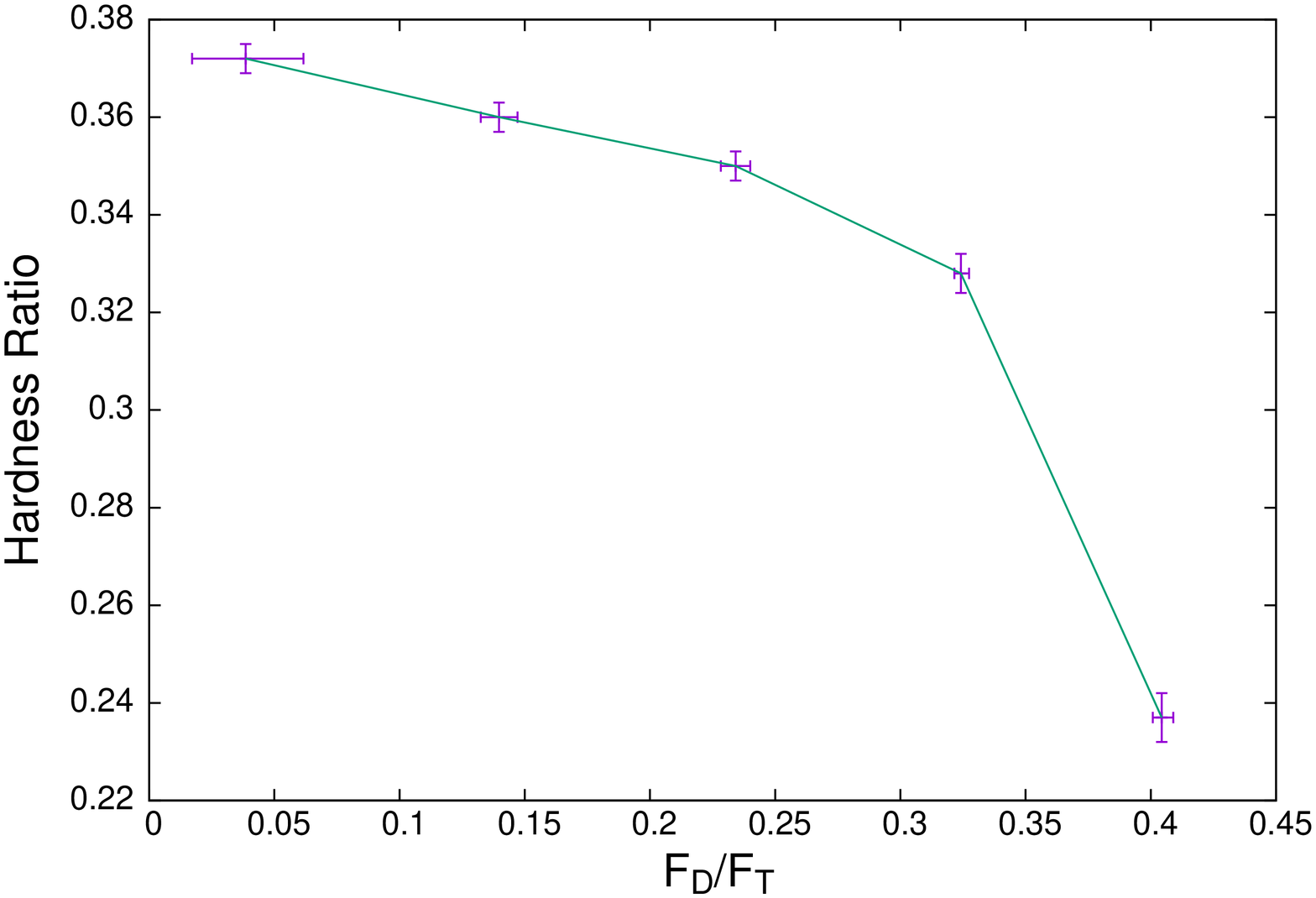}
 \centering \includegraphics[width=0.30\textwidth,angle=-90]{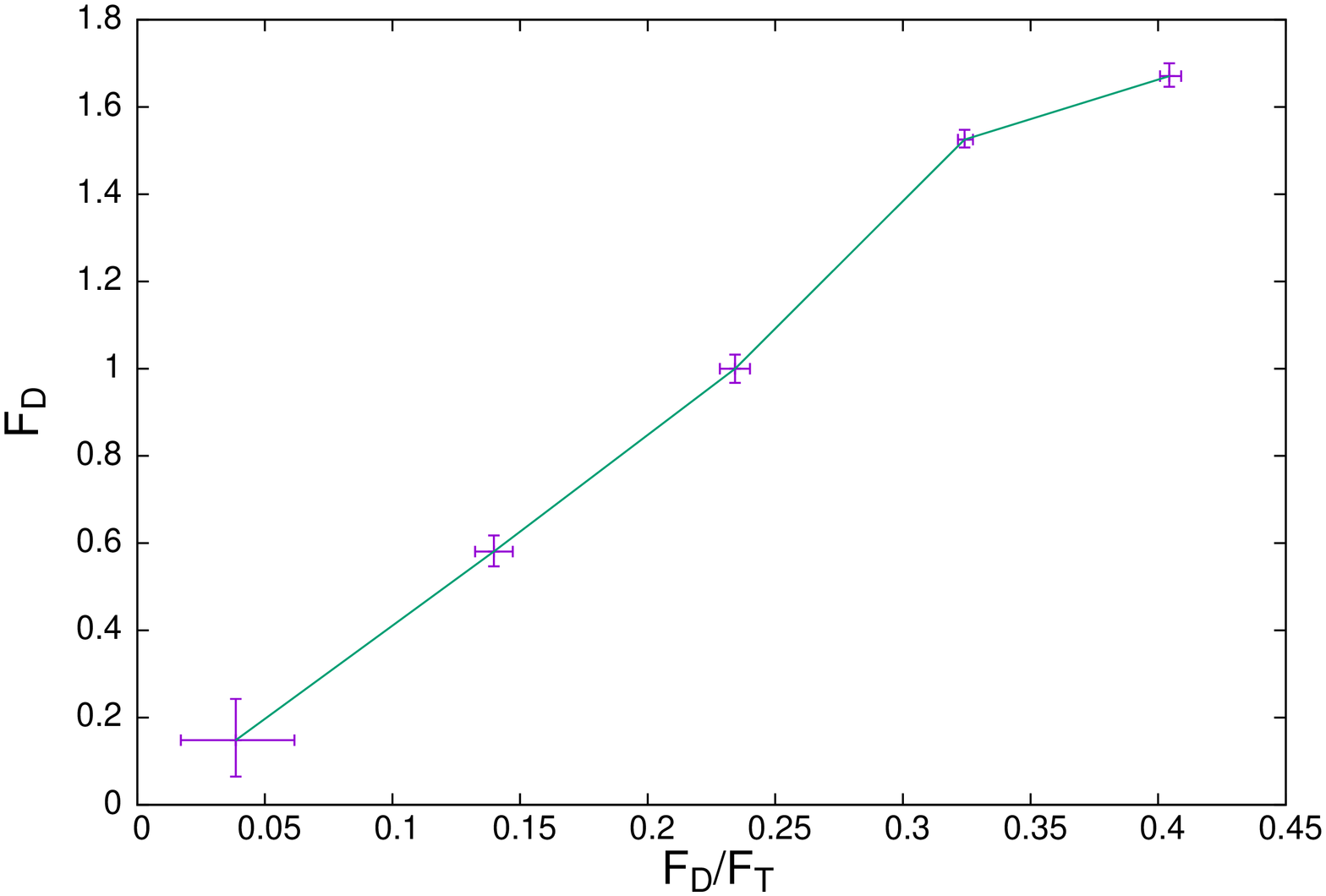}
\caption{Left Panel: Plot between hardness ratio as defined in figure \ref{msxxfig2} versus the disk flux ratio $F_D/F_T$,
 Right Panel: Variation of the bolometric disk flux $F_D$ with disk flux ratio $F_D/F_T$.}
\label{msxxfig4}
\end{figure*}

 \begin{figure}
\centering \includegraphics[width=0.40\textwidth,angle=-90]{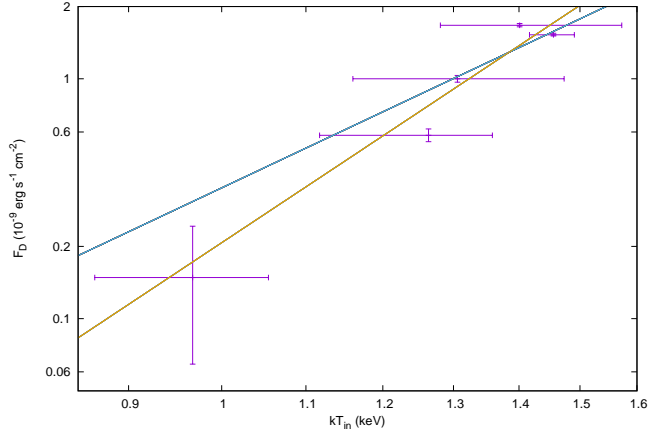}
\caption{Variation of the inner disk temperature $kT_{in}$ with the disk flux $F_D$. The best fit curve $F \propto T^{5.5}$ is marked in yellow color. Also
  shown is the relation $F \propto T^{4}$ in cyan colour. The data seems to be not consistent with $F \propto T^{4}$ i.e. a constant inner disk radius.}  
\label{msxxfig5}
\end{figure}

\section{Spectral analysis and results}
Spectral fitting was undertaken for each of the six different locations along the Z-track. For each location only that time-segment which was strictly simultaneous with an SXT observation was taken into consideration. This is illustrated in Figure \ref{msxxfig1}, where the vertical lines mark the time for which spectra  were analyzed. The exception was for the upper normal branch (UNB) where there are no corresponding SXT observation and thus for this segment, only the LAXPC data was considered.

We first fitted a two component model consisting of a Comptonized component and a blackbody one represented by the XSPEC routine ``nthcomp'' and ``bbodyrad'' respectively. The absorption was taken into account by the XSPEC routine ``tbabs''. The black body was assumed to be the seed photon source of the Comptonizing medium i.e. the temperature of the blackbody was tied to the input seed photon temperature during fitting and the shape of the seed photon was taken to be blackbody. While the fit produced reasonable $\chi^2$, the normalization of the blackbody component was extremely high. Since the normalization of the routine ``bbodyrad'' represents the radius of the black body sphere ($N_{bbodyrad} = R_{km}^2/D_{10 kpc}^2 $), we found that for a distance of $D = 9$ kpc, the radius turned out to have an  unphysically large value of $\sim$2000 km or more. This is true for all the segments considered with SXT data.

Our favoured model is the one where the blackbody is replaced  by a multi-colour disk blackbody(XSPEC routine: ``diskbb'') and the form of the seed photon for the Comptonization is taken to be such a disk emission.  The best fit parameters for the model are tabulated in Table \ref{msxxtable1} and a typical unfolded spectra with residuals is shown in Figure \ref{msxxfig3}.
The parameters are the normalization of the disk component $N_{dbb}$, the inner disk temperature $kT_{in}$, the
temperature of the Comptonizing medium $kT_e$ and the photon index $\Gamma$ of the Comptonized component. The disk normalization
is related to inner disk radius (in kms)  $R_{km}$ by  $=(R_{km}/D_{10kpc})^2{\ast}cos$ $i$,
where $D_{10kpc}$ is the distance to the source in units of 10 kpc and $i$ is the inclination angle of the disk.
The best fit values of the normalization are $\sim 180$, indicating  reasonable values of the inner disk radius
of $\sim 20$ kms. Note that the disk normalization and the absorption column density could not be constrained
for the UNB due to absence of SXT data and for that spectrum the normalization and column density was fixed. Hence the UNB
has been excluded from further analysis.

  We note that there have been several works where more complex models have been
  invoked to fit the spectra from such sources \citep[e.g.]{2018homan,Lin09}. In general, these models have three components instead of the two used here and hence
  we tested the possibility of including another component.
Specifically, we added an additional blackbody emission in the
model consisting of disk emission and thermal Comptonization. We find
that for all the spectra considered the decrease in $\chi^2$ is less than
3 and hence the data does not justify the inclusion of a third component.
We have also tried other variations of the three component model, such as assuming that the seed photon
for the thermal component is due to the blackbody instead of the disk, but
again find no significant improvement in the fitting. We note that despite
having a broader energy coverage than RXTE, we are unable to justify the
use of more complex three component models as used by \citet{Lin09}. We suspect that
the 3\% systematic that is included for the LAXPC and SXT spectra, does not
allow for more complex model to be constrained and discuss the implication of
this in the last section.

The unabsorbed bolometric fluxes (in the 0.001-100 keV range) was computed using the XSPEC function ``cflux'' for the
total and for the disk component. The total flux and the ratio of the disk flux to the total
and (which we call the disk flux ratio) are tabulated in Table \ref{msxxtable1}. While the total flux varies non-uniformly,
the disk flux ratio has a remarkable monotonous behaviour as increasing along the horizontal and the normal branch in
a uniform manner. This is illustrated more clearly in the left panel of Figure \ref{msxxfig4},
where the hardness ratio (8-20 keV by 3-8 keV) of the segment is plotted against the disk flux ratio. Thus, the location of the source in the horizontal and normal branches is determined by the value of the disk flux ratio making it the single parameter determining the spectral state of the system. We also note that the disk flux itself is found to be proportional to the disk flux ratio as shown in the right panel of Figure  \ref{msxxfig4}. The disk flux itself maybe the main driver behind the variability, a point which is discussed in the last section. 
There is no significant qualitative variation in the Comptonization parameters 
namely the electron temperature, $kT_e$  and photon index, $\Gamma$. 

The inner disk radius shows variability, which is more clearly exhibited by plotting the inner disk temperature versus
the disk flux as shown in Figure \ref{msxxfig5}. For a constant inner radius the flux should scale as $T_{in}^4$, but instead
the flux scales more steeply as $T_{in}^{5.5}$. It may also be possible and perhaps more likely that the colour factor
of the disk is changing (perhaps due to irradiation) causing this steeper variation, while the inner disk radius
is more or less constant. On the other hand, this may also indicate that the model used here is simple and a more complex spectral model
  may be needed.

   \begin{table*}
   \centering
   \begin{center}
   \begin{tabular} {c c c c c c} 
\hline 

 Location & $^{*}NB$ & HA & HB-1 & HB-2 & HB-3   \\ 
\hline
${\nu_1}$ & 0 & 0 & 0 & 0 & 0 \\ 
$\sigma_1$&$2.53_{-0.47}^{+0.52}$ & $3.47_{-0.55}^{+0.57}$  & $18.64_{-2.52}^{+1.35}$  &$13.30_{-1.18}^{+1.33}$ &  $10.08_{-1.16}^{+1.32}$\\
$N_1(\times10^{-4})$ & $0.23_{-0.1}^{+0.1}$   &  $2.39_{-0.37}^{+0.36}$ & $13.90_{-4.1}^{+4.9}$  &  $ 21.29_{-0.97}^{+0.96}$  &  $26.30_{-1.5}^{+1.5}$\\

${\nu_2}$ & $5.99_{-0.65}^{+0.70}$ & $5.35_{-1.44}^{+1.28}$  & - & -  & - \\
$\sigma_2$&   $4.65_{-1.76}^{+2.46}$   &  $19.62_{-0.38}^{+0.40}$  &  -  & - & -\\
$N_2(\times10^{-4})$ & $2.28_{-0.09}^{+0.67}$ & $5.38_{-0.43}^{+0.36}$  & -  & - & -\\

${\nu_3}$ & $49.62_{-0.57}^{+0.48}$ & $39.53_{-0.16}^{+0.18}$   & $37.22_{-0.17}^{+0.17}$   & $34.20_{-0.17}^{+0.16}$  & $30.19_{-0.10}^{+0.09}$ \\
$\sigma_3$& $6.97_{-1.79}^{+2.37}$  & $19.36_{-1.1}^{+0.64}$  &  $8.99_{-0.59}^{+0.64}$ &  $7.66_{-0.70}^{+0.80}$  &  $4.52_{-0.15}^{+0.15}$ \\
$N_3(\times10^{-4})$& $0.81_{-0.31}^{+0.57}$ & $6.85_{-0.13}^{+0.14}$   & $7.97_{-0.39}^{+0.28}$  & $ 12.07_{-0.74}^{+0.77}$ & $13.67_{-1.1}^{+1.1}$ \\

${\nu_4}$ & - & - & $74.22_{-5.82}^{+4.67}$ & $68.31_{-7.08}^{+7.18}$ & $58.41_{-5.11}^{+4.59}$ \\
$\sigma_4$& - &  - &  $15.39_{-7.42}^{+2.57}$  & $15.32_{-4.51}^{+4.68}$   &$18.04_{-6.28}^{+1.96}$ \\
$N_4(\times10^{-4})$& -  & - &   $6.32_{-0.27}^{+0.27}$  & $11.20_{-0.79}^{+0.83}$  & $20.60_{-0.71}^{+0.72}$ \\

    \hline
    \end{tabular}
    \end{center}
  \caption {${\nu_1}$, ${\nu_2}$, ${\nu_3}$ \& ${\nu_4}$ are the frequency in Hz, $\sigma_1$, $\sigma_2$, $\sigma_3$ \& $\sigma_4$ are the frequency width at half maximum (FWHM) line width in keV, and  $N_1$, $N_2$, $N_3$ \& $N_4$ are the normalisations of the lorentzians. $(^*)$ indicates the QPO is for full NB region, since it was not detected individually in LNB and UNB due to poor statics.} 
 \label{msxxtable2}
    \end{table*}

 \begin{figure*}
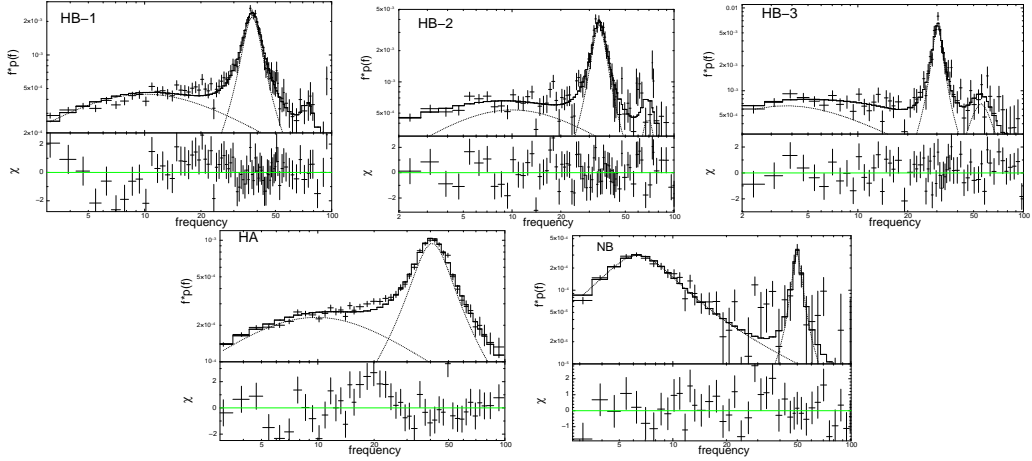

\centering \includegraphics[width=0.20\textwidth,angle=-90]{86-94-3-20-pds.ps}
\centering \includegraphics[width=0.20\textwidth,angle=-90]{81-86-3-20-pds.ps}
\centering \includegraphics[width=0.20\textwidth,angle=-90]{73-81-3-20-pds.ps}
\centering \includegraphics[width=0.20\textwidth,angle=-90]{ha-7664-7665-3-20.ps}
\centering \includegraphics[width=0.20\textwidth,angle=-90]{nb-3-20.ps}
\caption{AstroSat Power spectra of GX 5-1 for each of the five locations of the Z-track in the energy range 3-20 keV. The power spectra has been fitted by three lorentzian components and the best fit parameters are listed in Table \ref{msxxtable2}.}
\label{msxxfig6}
\end{figure*}

  \begin{figure*}
\centering \includegraphics[width=0.17\textwidth,angle=-90]{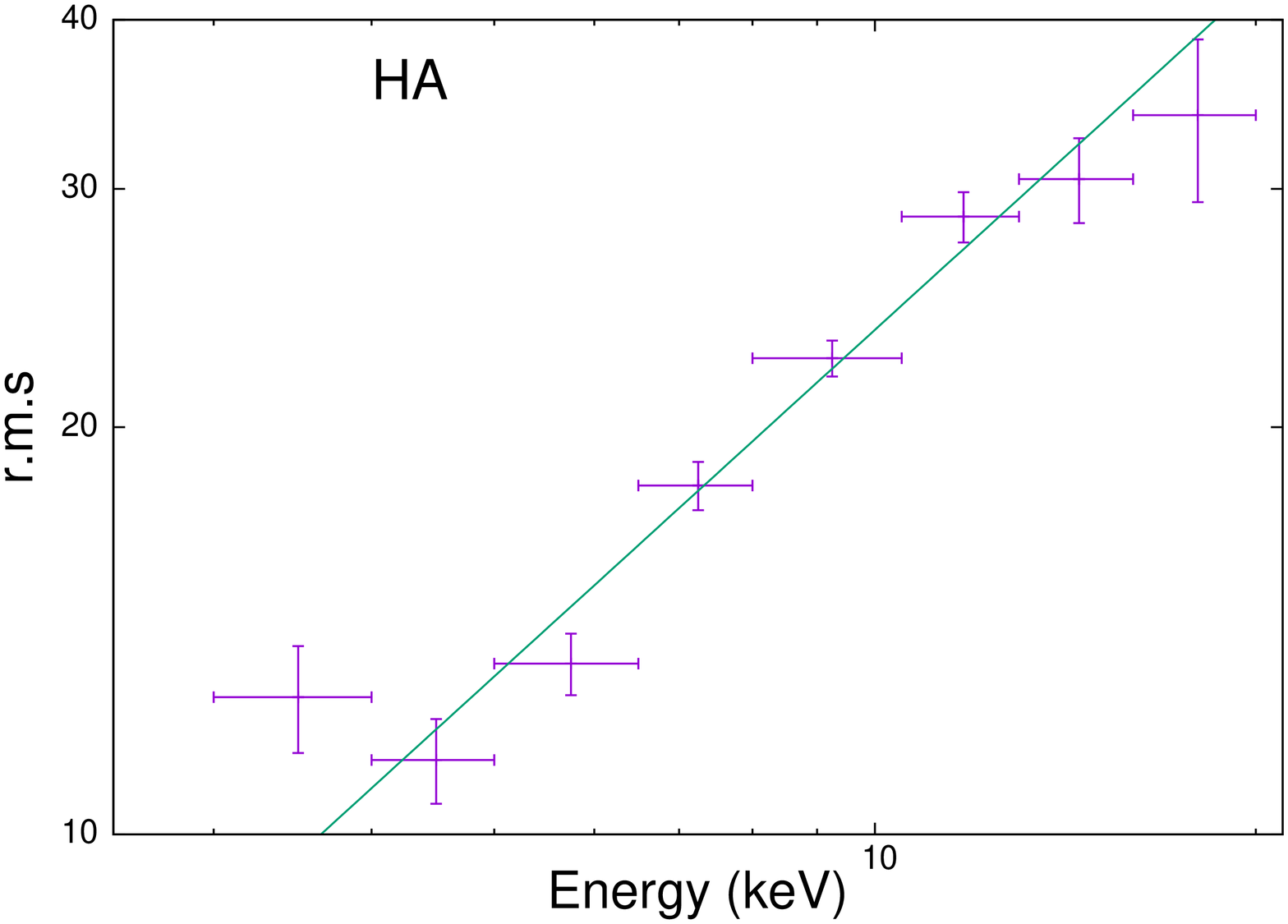}
\centering \includegraphics[width=0.17\textwidth,angle=-90]{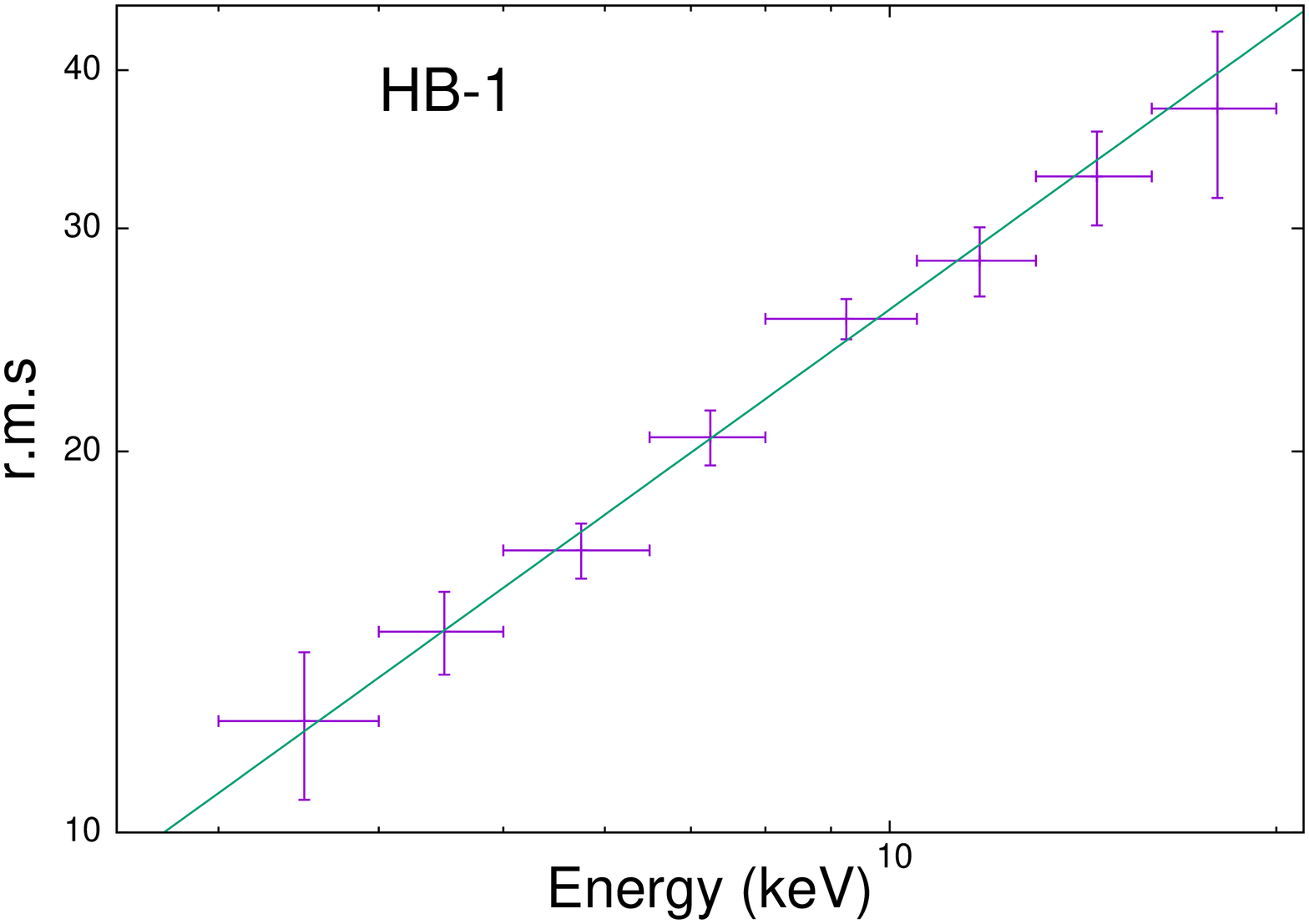}
\centering \includegraphics[width=0.17\textwidth,angle=-90]{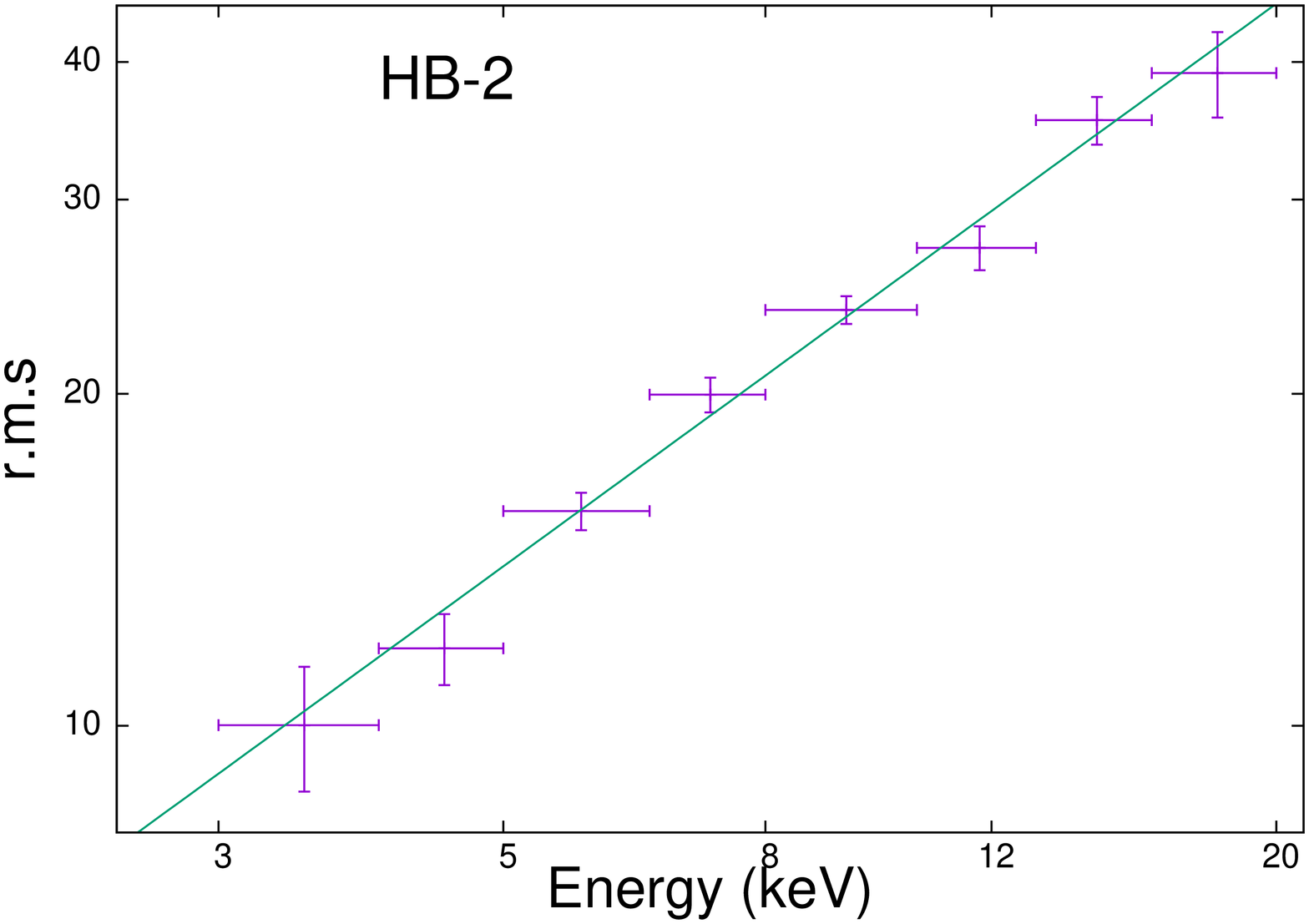}
\centering \includegraphics[width=0.17\textwidth,angle=-90]{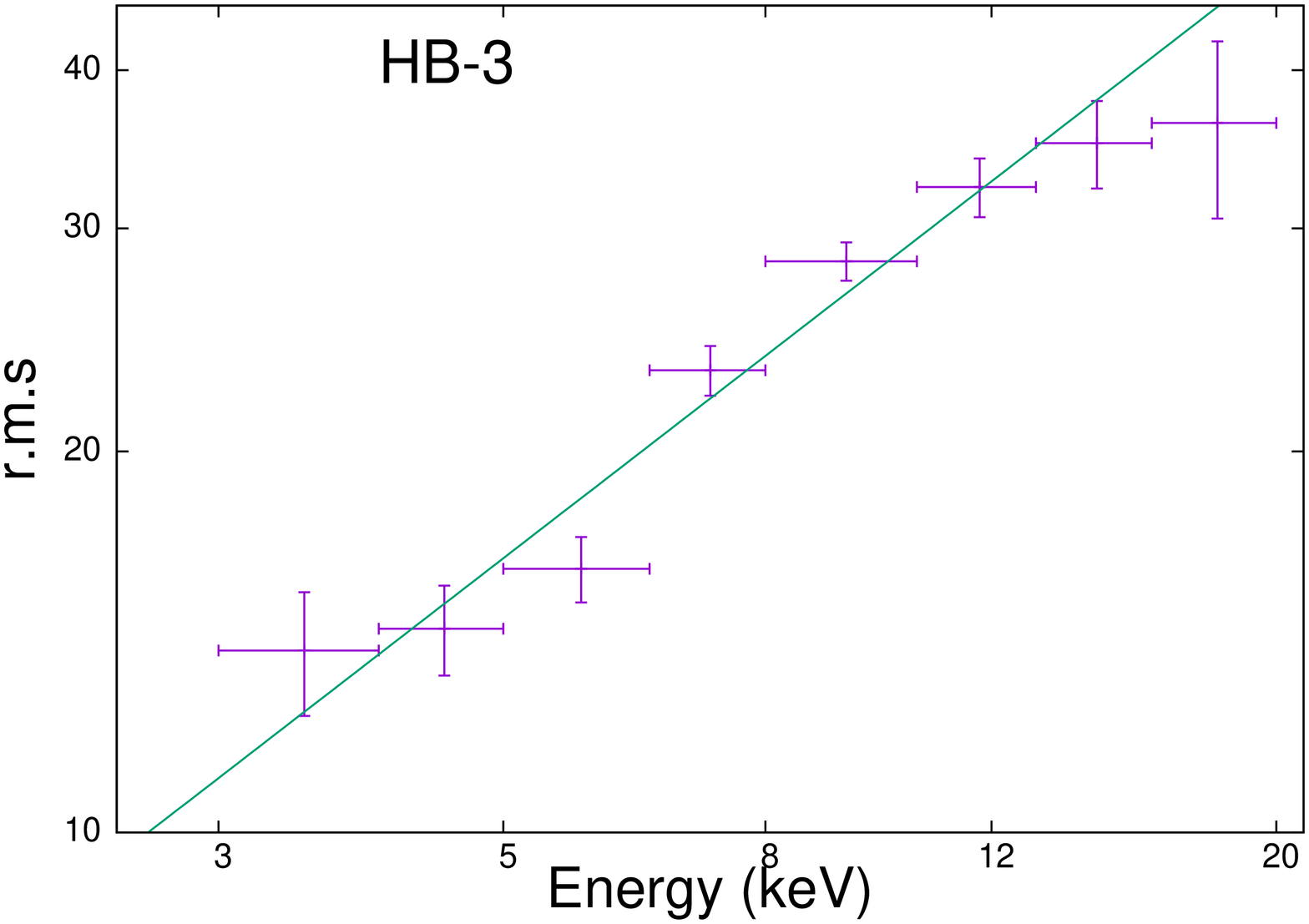}
\caption{The root mean square (r.m.s) of the QPO versus photon energy for three different regions of the horizontal branch
and the Hard Apex. The r.m.s roughly increase with energy $\propto E^{0.8}$.}
\label{msxxfig7}
\end{figure*}

\begin{figure*}
\centering \includegraphics[width=0.17\textwidth,angle=-90]{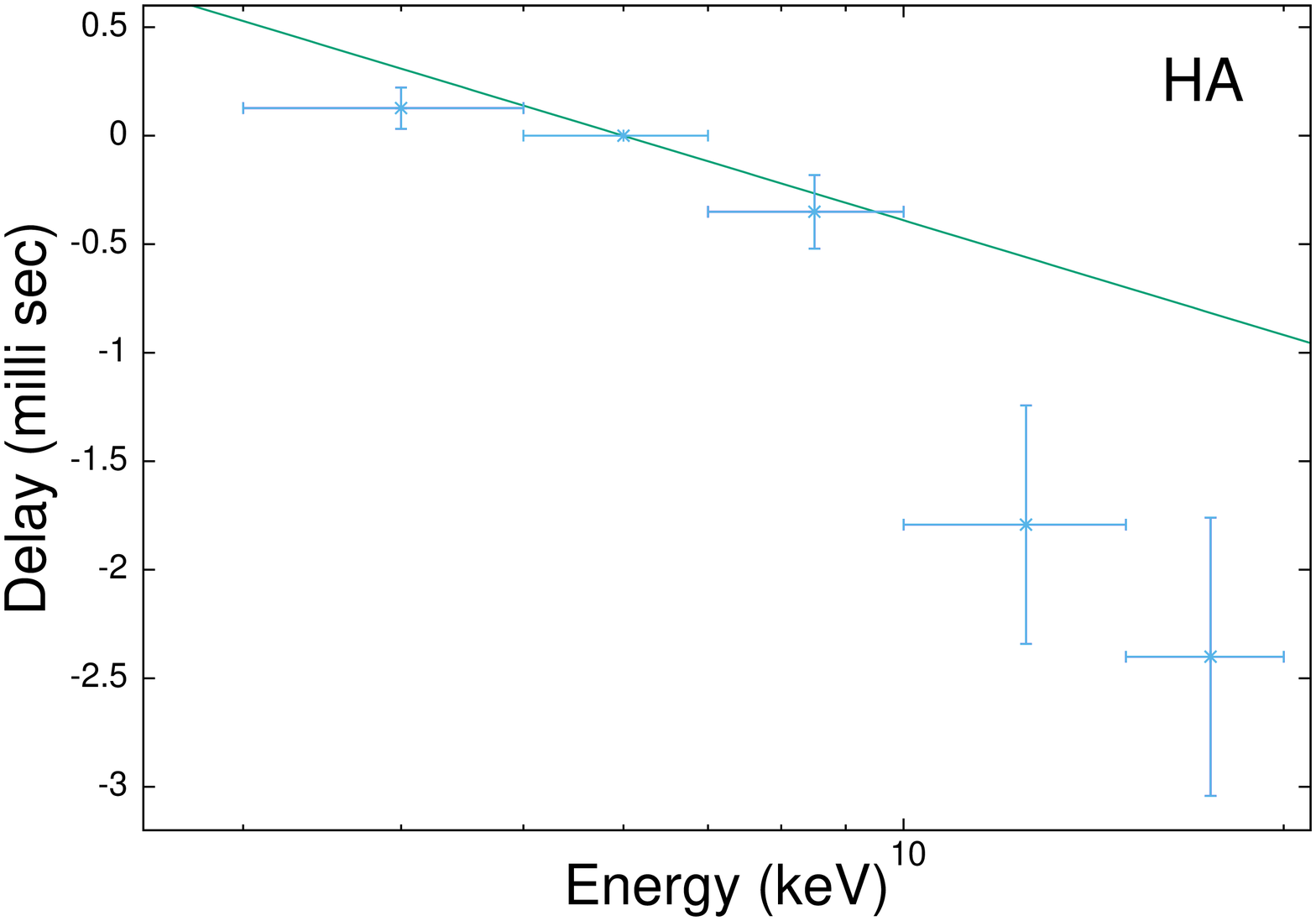}
\centering \includegraphics[width=0.17\textwidth,angle=-90]{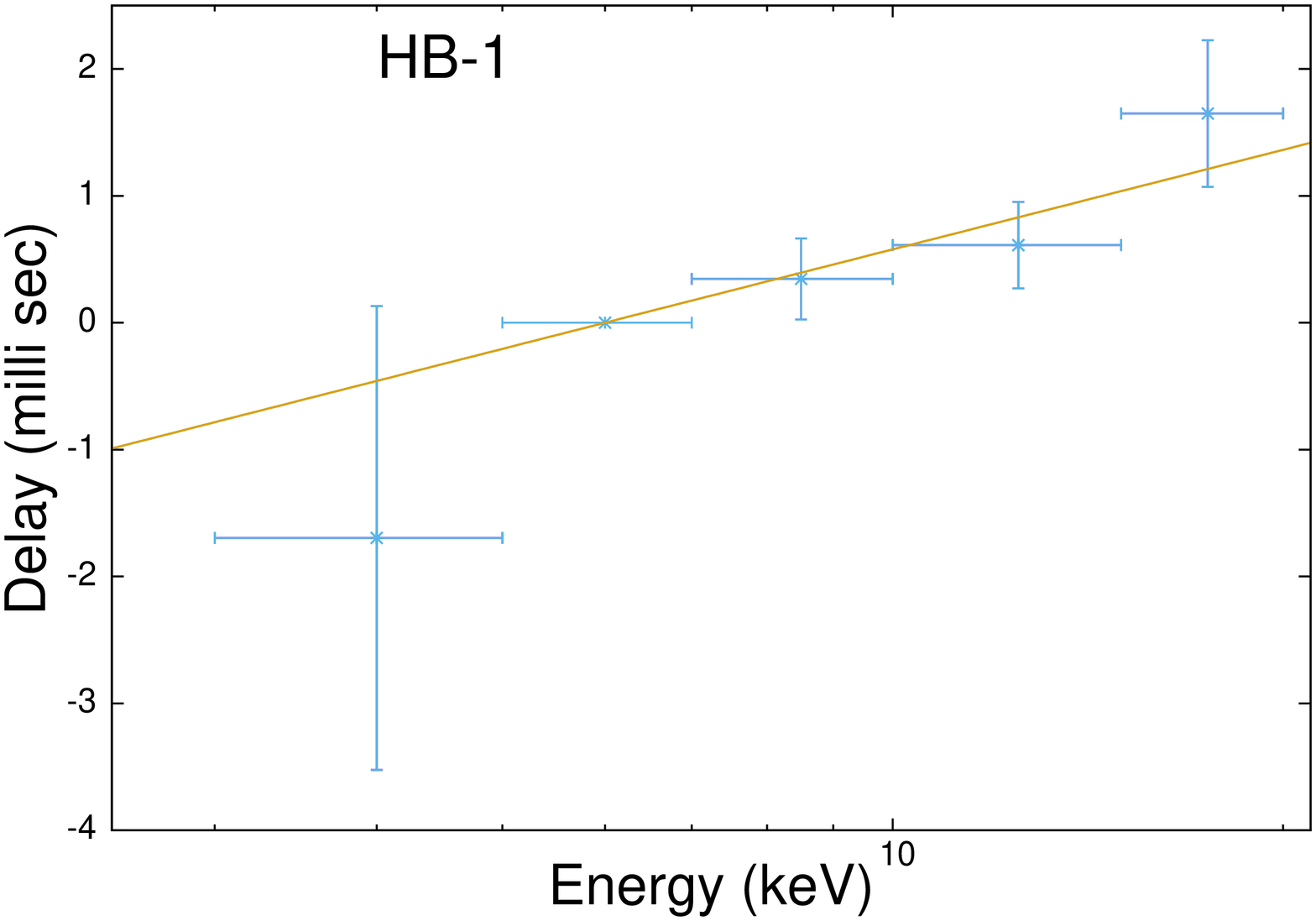}
\centering \includegraphics[width=0.17\textwidth,angle=-90]{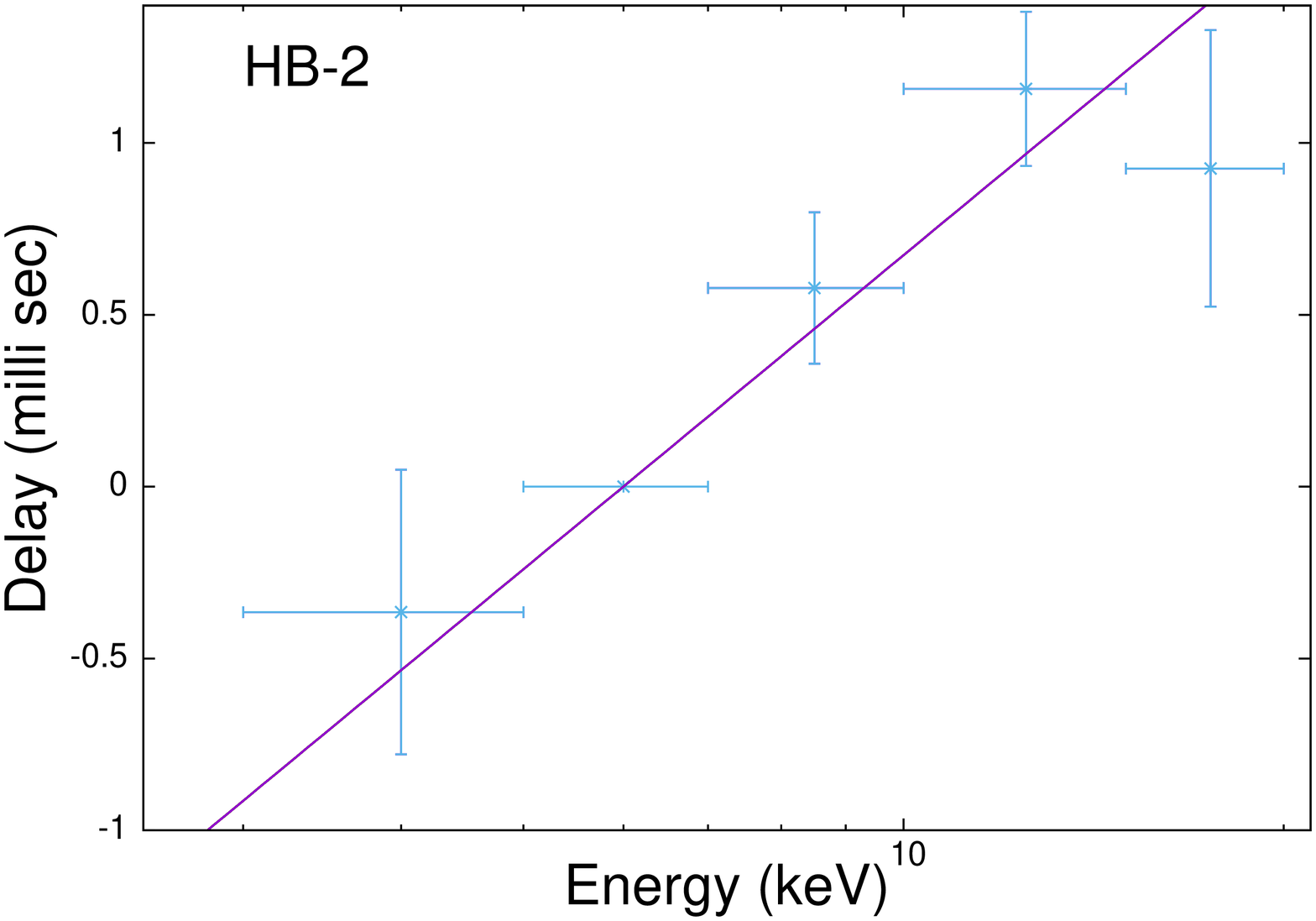}
\centering \includegraphics[width=0.17\textwidth,angle=-90]{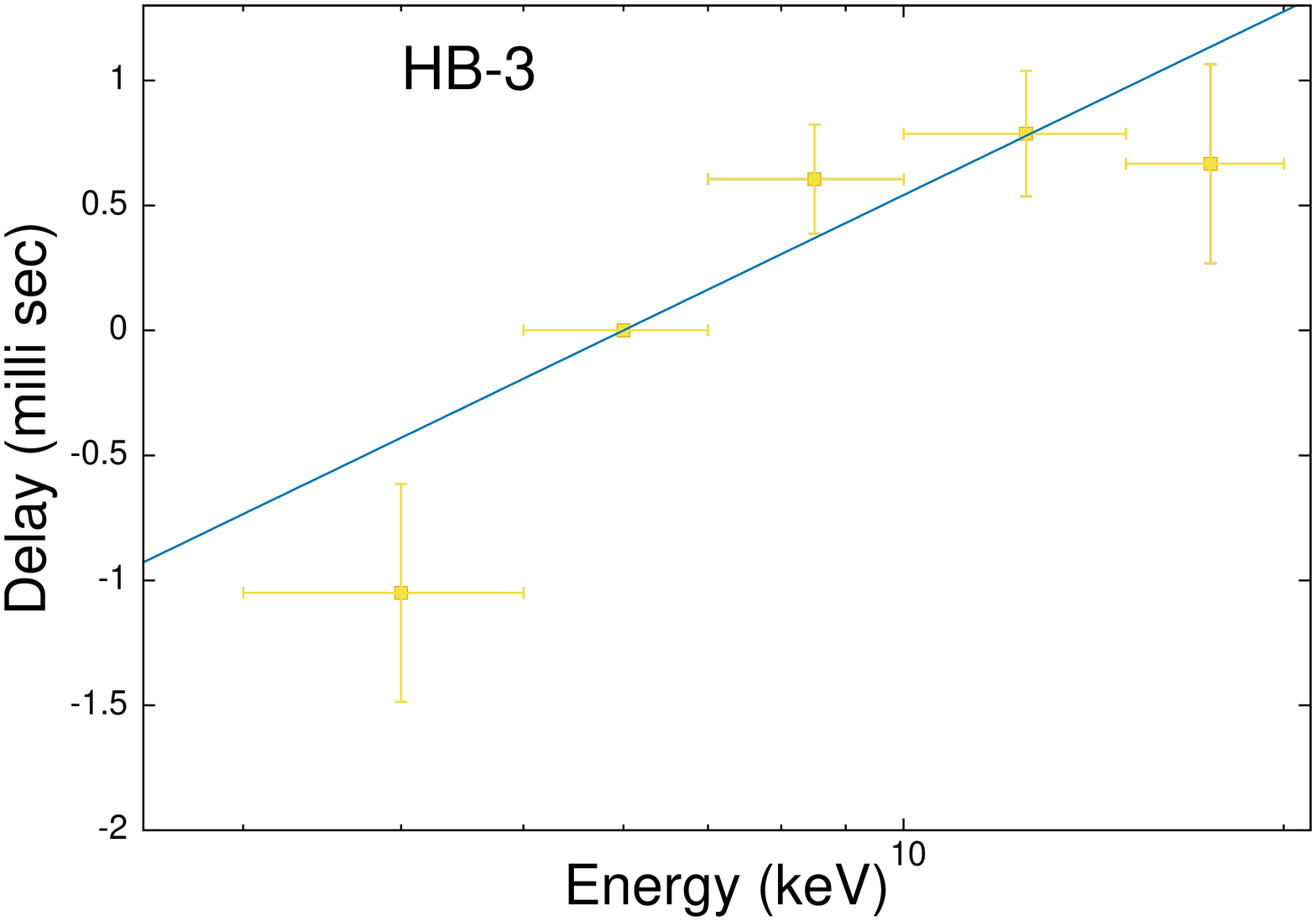}
\caption{Time Lag at the QPO frequency with respect to  5--7 keV band versus energy for three different regions of the horizontal branch and the Hard Apex. While the time-lags seem to be hard for the horizontal branch, it seems to be soft for the Hard apex. However, the statistics are not good. Note that the time-lags are constrained to be less than a few milli-seconds.  }
\label{msxxfig8}
\end{figure*}

\begin{figure}
\centering \includegraphics[width=0.30\textwidth,angle=-90]{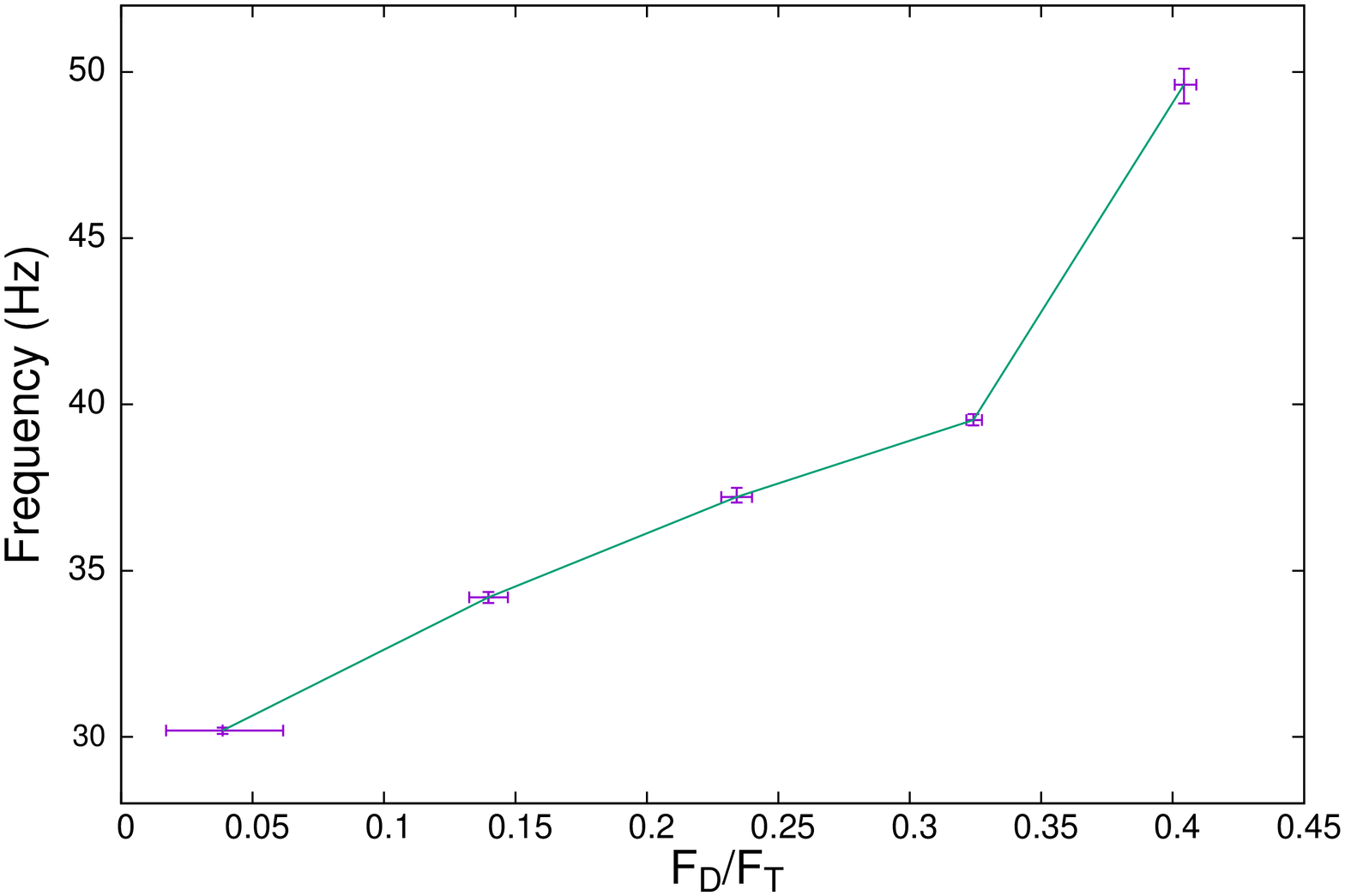}
\centering \includegraphics[width=0.30\textwidth,angle=-90]{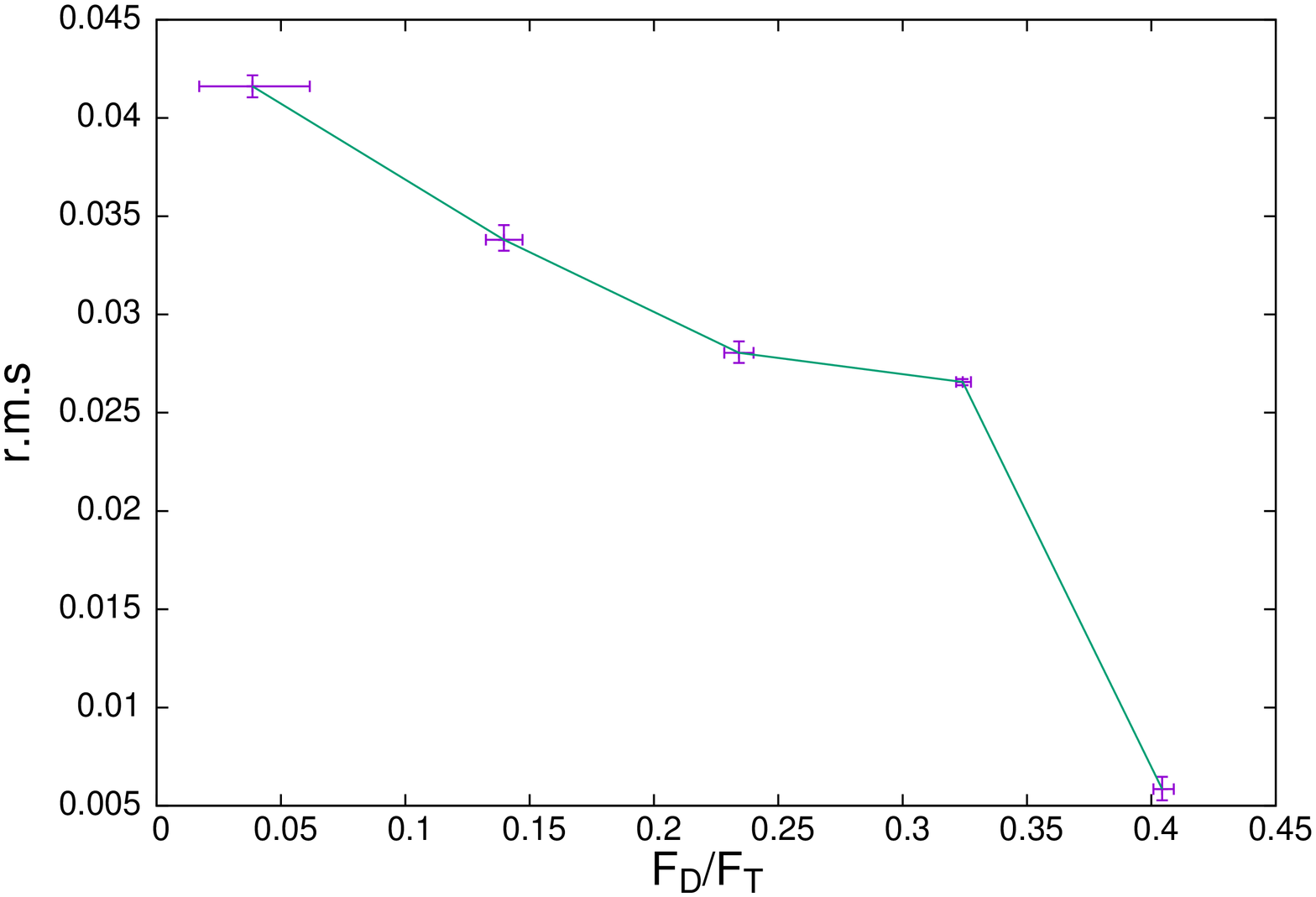}
\caption{The variation of the QPO frequency and rms as a function of the disk flux ratio $F_D/F_T$. While the frequency increases with the ratio, the r.m.s decreases.}
\label{msxxfig9}
\end{figure}

\section{Timing analysis}
 Power density spectra (PDS) were computed for each of the Z-track sections in the 3-20 keV band. All three LAXPC counters were
 used in the timing analysis and the expected dead time corrected Poisson noise level was subtracted from each power spectra
 which were normalized such that the integration of the power spectra gave the fractional root mean square (r.m.s) variation.
 The top panel of Figure \ref{msxxfig6} shows the power spectra for the three sections of the horizontal branch. A clear QPO
 is detected which dominates the power spectra. The  power spectra were fitted using three Lorentzians corresponding to
 the QPO, its harmonic and a zero centered low frequency broad band noise for the horizontal Branch sections. For the Hard Apex
and normal branch, the harmonic is not seen while another broad feature at $\sim 5$ Hz is required by the data. 
The best fit values of the fitting are tabulated in Table
 \ref{msxxtable2}.  For the normal Branch, the QPO was not detected in the individual  two sections, due to the shorter exposure time and hence we combined the two section into one and the power spectrum  is shown in the right bottom panel of Figure \ref{msxxfig6}. We note that the low frequency component has significantly  increased during the normal branch. The QPO frequency increases monotonically along the horizontal to the normal branch  as can be seen in top panel of Figure \ref{msxxfig9}, where the frequency is plotted against disk flux ratio. However, its fractional r.m.s decreases (bottom panel of Figure \ref{msxxfig9}).
 
 The Event mode data of LAXPC allows for computing the temporal properties in several energy bins. Power spectra were 
 analysed for different energy bins and the fractional r.m.s obtained from fitting a Lorentzian to the QPO was estimated
 for the different energy bins. Figure \ref{msxxfig7} shows the variation of the fractional r.m.s with energy for the three
 sections of the horizontal Branch and the Hard Apex. The statistics for the normal branch was not good enough to do such
 an analysis. The variation of r.m.s with energy is similar for all the sections. It increases with energy and
 its functional dependence can be described as $\propto$ $E^{0.8}$.
 
 LAXPC data also allows the computation of the time-lag as a function of energy. To compute the time-lag the
 Fourier transforms were undertaken with a frequency bin equal to the FWHM of the QPO and lags were computed
 at the QPO frequency with respect to the reference energy band of 5-7 keV. Figure \ref{msxxfig8} shows the
 time-lag as a function of energy for the different sections of the horizontal branch and the Hard Apex.
 It seems that the horizontal branch segments show positive hard time-lag  (i.e. the high energy photons are delayed compared to the soft) while for the Hard Apex it is the opposite. The positive time-lag seen in the HB does not seem  not seem to be
   correlated with the
 position in the Z-track.  However, since the statistics are not good this may not be significant and we caution against over interpretation.  What is clear is that the magnitude of the time-lags are less than a few milli-seconds.

\section{Summary and discussion}

We present the analysis of an AstroSAT observation of the Z-track source GX 5-1. We used the strictly simultaneous SXT and LAXPC data to track the spectrum of a neutron star source as it moves along its
Hardness intensity diagram (HID) and at the same time quantify the rapid temporal variability of the source. During the observation, GX 5-1 tracked out the normal and horizontal branches in its HID which was divided into six segments and analyzed separately.

The primary results are:-

\noindent $\bullet$ The broad band SXT/LAXPC spectra from 0.8-20 keV was fitted with an absorbed black body and thermal component, but the radius of the blackbody turned out to be unphysically large $\sim$ 2000 kms. On the other hand an absorbed  disk emission and a thermal Comptonized component fitted the data well and the inner disk radii turned out to be reasonable i.e. $\sim 20$ kms. Thus the latter model was preferred.

\noindent $\bullet$ The disk flux ratio defined as the ratio of the bolometric unabsorbed disk flux to the total unabsorbed flux, decreased monotonically along the horizontal branch to the normal one, indicating that this may be the primary driver which determines the location of the source in the Z diagram. However, we note that the disk flux itself is correlated with the disk flux ratio. The disk flux varies as $T_{in}^{5.5}$ indicative of a varying inner radius or changes in the colour factor.

\noindent $\bullet$ The power spectra for all sections of  the Z-track shows a QPO at $\sim 50$ Hz. The QPO is dominant in the horizontal branch while its strength decreases in the normal one, where the power spectrum also has a pronounced low frequency component. The QPO frequency increases,  while its r.m.s decreases with disk flux ratio. The r.m.s of the QPO increases with energy ($\propto$ $E^{0.8}$) for all sections of the Z-track. The time lags between different energy bins are constrained to be less than a few milliseconds.

One of the important clues regarding the temporal behaviour of these systems is to identify the physical parameter (or at least the spectral parameter) that determines the source's position in the HID diagram. Our results suggest that this may be related to the disk flux ratio. Thus the difference in hardness for horizontal and normal branches (at the same intensity level) is the contribution of the disk to the total flux. It is interesting to note that there is also a correlation between the disk flux and the disk flux ratio, perhaps indicating that the disk flux itself is the primary driver. If that is the case, and if we assume that the inner disk radii do not vary much, then the disk flux would be proportional to the accretion rate, and one could get back the physically attractive proposition, that the accretion rate itself determines the state of the system. However, this implies that the total flux, and in particular, the flux of the thermal Comptonization component does not scale with the accretion rate. For a neutron star system with a surface, one expects the total flux to be proportional to the accretion rate since unlike a black hole, the accretion efficiency cannot be small due to advection. On the other hand, if a significant amount of energy or matter is given out to a jet or wind, then the radiative flux may indeed not be proportional to the accretion rate. Such a speculation needs to be followed up with  evidence for jet or wind activity perhaps in non X-ray wavebands.

  The spectral analysis undertaken in this work seems to imply that there
  are only two spectral components, namely the disk emission and a thermal
  component. This is in contrast to other works \citep[e.g]{Lin09}, where a
  third blackbody component has been included even though the energy band
  used in this work $0.8-20$ keV is broader than that of the RXTE analysis
  $> 3$ keV. Note that for spectral analysis
  of RXTE data, the column density was taken to be fixed, while here the presence of low energy data allows us to constrain it.  It is not clear, whether the broader energy band constrains the
  fitting to have lesser number of components or if the systematics used
  for the fitting (3\%) is too large to discriminate between complex models.
  This may be the reason why in this analysis the disk flux does not scale
  as temperature to the fourth power, while the analysis of \citet{Lin09} showed
  such a scaling for the three component model. Low energy spectral coverage
  from data from other instruments like NICER will be useful to determine whether the spectra is a simple or complex one.

The rapid temporal behaviour of the source shows a strong QPO at $\sim 50$ Hz whose frequency increases, while its r.m.s decreases with the disk flux ratio. Energy dependent analysis of the QPO reveals that the r.m.s is an increasing function of energy and time-lags between high and low energy photons are less than a few milliseconds. This implies that the QPO originates
in the thermal Comptonized component and it weakens when the component is less dominant as has been shown earlier using RXTE data \citep{2011srirama}. Detailed radiative modelling of the
these energy dependent properties will shed light on the driver and the geometry of the system.  
The QPO frequency should be associated with some characteristic time-scale of the system, which in turn may be related to come characteristic radius. Since the time-scale should decrease with radius, the frequency should be inversely proportional to such a radius. It is tempting to associate this radius
  with the inner disk one. While the inner disk radius estimated in the
  spectral fitting maybe subjected to several uncertainties like the spectral
  model used, color factor etc, it is interesting to see the correlation between
  the frequency and disk flux ratio. This is qualitatively expected if the
disk flux ratio is inversely related to the inner disk radius. 

The broad band spectral coverage provided by SXT/LAXPC along with the rapid timing capabilities of the LAXPC has provided
a unique opportunity to study these neutron star systems. Note for example, that when there was no simultaneous SXT spectrum for the upper section of the normal Branch, the disk normalization and absorption column density was not constrained, which highlights the importance of SXT for such analysis. Indeed, other observations by AstroSat of GX 5-1 and other such sources will provide insight into their temporal behaviour.

\section{Acknowledgments}

We thanks to members of LAXPC instrument team for their contribution to the development of the LAXPC instrument. We also acknowledge contributions of the AstroSat project team at ISAC. This paper makes use of data from the AstroSat mission of the Indian Space Research Organisation (ISRO), archived at the Indian Space Science Data Centre (ISSDC). We also acknowledge
support of LAXPC POC as well as  SXT POC at TIFR.

\bibliographystyle{raa}
\bibliography{cm_ypbhulla_gx5_RM.bib}
\end{document}